\documentclass[preprint,12pt]{aastex}
\begin{document}
%\begin{CJK*}{Bg5}{bsmi}
%
\title{The Magnetic Field Morphology of the Class 0 Protostar L1157-mm}
%
%=================================================================
% authors
%=================================================================
%
\author{Ian W. Stephens\altaffilmark{1}, Leslie W. Looney\altaffilmark{1,2}, Woojin Kwon\altaffilmark{1,3}, Charles L. H. Hull\altaffilmark{4}, Richard L. Plambeck\altaffilmark{4}, Richard M. Crutcher\altaffilmark{1}, Nicholas Chapman\altaffilmark{5}, Giles Novak\altaffilmark{5}, Jacqueline Davidson\altaffilmark{6}, John E. Vaillancourt\altaffilmark{7}, Hiroko Shinnaga\altaffilmark{8}, Tristan Matthews\altaffilmark{5}}
%\end{CJK*} %End the Chinese part
\altaffiltext{1}{\itshape Department of Astronomy, University of Illinois, 1002 West Green Street, Urbana, IL 61801; stephen6@illinois.edu}
\altaffiltext{2}{\itshape National Radio Astronomy Observatory, 520 Edgemont Road, Charlottesville, VA 22903, USA.National Radio Astronomy Observatory, 520 Edgemont Road, Charlottesville, VA 22903, USA.}
\altaffiltext{3}{\itshape SRON Netherlands Institute for Space Research, Landleven 12, 9747 AD Groningen, The Netherlands}
\altaffiltext{4}{\itshape Astronomy Department and Radio Astronomy Laboratory, 601 Campbell Hall, University of California, Berkeley, CA 94720, USA}
\altaffiltext{5}{\itshape Center for Interdisciplinary Exploration and Research in Astrophysics (CIERA) \& Dept. of Physics \& Astronomy, Northwestern University, 2145 Sheridan Road, Evanston, IL 60208}
\altaffiltext{6}{\itshape School of Physics, University of Western Australia, 35 Stirling Hwy, Crawley, WA 6009, Australia}
\altaffiltext{7}{SOFIA Science Center, Universities Space Research Association, NASA Ames Research Center, MS 232-11, Moffett Field, CA 94035-0001, USA}
\altaffiltext{8}{\itshape Subaru Telescope, National Astronomical Observatory of Japan, 650 North A'ohoku Place, Hilo, HI 96720, USA}

\begin{abstract}

We present the first detection of polarization around the Class 0 low-mass protostar L1157-mm at two different wavelengths. We show polarimetric maps at large scales (10$\arcsec$ resolution at 350 $\mu$m) from the SHARC-II Polarimeter and at smaller scales (1.2$\arcsec$-4.5$\arcsec$ at 1.3~mm) from the Combined Array for Research in Millimeter-wave Astronomy (CARMA). The observations are consistent with each other and show inferred magnetic field lines aligned with the outflow. The CARMA observations suggest a full hourglass magnetic field morphology centered about the core; this is only the second well-defined hourglass detected around a low-mass protostar to date. We apply two different methods to CARMA polarimetric observations to estimate the plane-of-sky magnetic field magnitude, finding values of 1.4 and 3.4~mG.

\end{abstract}

\interfootnotelinepenalty=10000

%=================================================================
\section{Introduction}  \label{sec:intro}
%=================================================================
Alignment of dust grains by magnetic fields causes the thermal emission from dust to be polarized \citep[e.g.,][]{hil88}. 
%Through polarization measurements of the dust continuum, the angle of the magnetic field with respect to the plane-of-sky can be ascertained. 
The angles of dust polarization measurements at millimeter and submillimeter wavelengths are generally considered to be perpendicular to the magnetic field due to a number of alignment mechanisms (particularly radiative torques, \citealt{laz07}), allowing for an estimate of the magnetic field direction in the plane of the sky. 

Magnetic pressure can support clouds against collapse, though other processes, e.g., ambipolar diffusion, can allow gravity to eventually overwhelm magnetic support. The importance of magnetic fields in star formation is not well understood (e.g., weak- versus strong-field models, \citealt{cru12}), but observations of the plane-of-sky magnetic field morphology can provide insight into the coupling of magnetic fields with cores, disks, and outflows. Although flux-freezing during gravitational collapse is expected to create an hourglass morphology in the magnetic field lines, there has only been one detection of a full hourglass around a low-mass protostar to date \citep[e.g.,][]{gir06}. However, several high-mass star formation regions have clear hourglass detections \citep[e.g.,][]{sch98,gir09,tan10}.

Large, submillimeter surveys of various star-forming clouds and cores have been performed with single-dish polarimeters \citep[e.g.,][]{mat09,dot10}. The SHARC-II Polarimeter \citep[SHARP,][]{li08} at the Caltech Submillimeter Observatory, is one of the newer submillimeter polarimeters and allows for dual-beam (simultaneous observations of horizontal and vertical polarization components) polarimetric measurements.

Previous millimeter-wavelength interferometric polarimetric maps have been produced with the Berkeley-Illinois-Maryland Association (BIMA), the Owens Valley Radio Observatory (OVRO), and the Submillimeter Array (SMA) \citep[e.g.,][]{gir06,kwo06,rao09,tan10}, but these observations only used single-polarization receivers modulated with a quarter-wave plate. The Combined Array for Research in Millimeter-wave Astronomy (CARMA) has recently installed 1~mm dual-polarization receivers; the first science observations in full-Stokes were made in 2011.% Unlike the SHARP observations, CARMA has the ability to measure left- and right-circular polarization.
% (i.e., Stokes $V$). 

In this letter we present polarimetric observations from SHARP and CARMA of the protostar L1157-mm in the dark cloud L1157. L1157-mm is a low-mass Class 0 source (i.e., the youngest of protostars) with a large bipolar outflow spanning about 5$\arcmin$ \citep{bac01} and a perpendicular 2$\arcmin$ flattened envelope \citep{loo07}. Though the distance to the source is a bit uncertain, with estimates ranging from 200 to 450~pc \citep[e.g.,][]{kun98}, we adopt a distance of 250~pc \citep{loo07}. 

These observations trace the magnetic field structure within the core and throughout the infall envelope (which, given the adopted distance, could extend to 30-40$\arcsec$) of L1157-mm. In this letter we will examine how the continuum and the magnetic field morphologies correlate with other structures in L1157, such as its outflow and flattened core, and will estimate the field strength. 

\section{Observations and Data Reduction} \label{sec:observations}
%=================================================================
With SHARP and CARMA we create polarimetric maps that show the fractional linear polarization, $P=\sqrt{Q^2+U^2}/I$ (where $I$, $Q$, and $U$ are Stokes parameters) and the angle with respect to the plane of the sky, $\theta$ (measured counterclockwise from north). When making uncertainty cuts based on polarization, we used polarized intensity, $P_I = IP$ rather than $P$. Due to the only positive nature of $P$ and $P_I$, de-biasing these values is necessary \citep[e.g.,][]{vai06} and was done for all observations.
%SHARP present large-scale submillimeter observations (350 $\mu$m), while CARMA observations probe smaller scales at radio wavelengths (1~mm).

\subsection{SHARP} \label{sec:SHARP_observations}
Polarimetric observations of L1157-mm were made with SHARP in September of 2008. These observations were at 350 $\mu$m (resolution of $\sim$10$\arcsec$) and were mosaicked in a chop/nod observing mode. The SHARP data reduction is discussed in \citet{cha13}. The total integration time on L1157 is about 15.8 hours; due to constant chopping, the integration time on the source is closer to 7 hours.

\subsection{CARMA} \label{sec:CARMA_observations}
%Beam-sizes
% C: sqrt(2.60573258856E-04*2.12379076402E-04)*3600 = 0.846883 PA: -8.41254959106E+01
% D: sqrt(5.94095385168E-04*5.43678179383E-04)*3600 = 2.04598 PA: -3.25446472168E+01
% E: sqrt(1.86807173304E-03*8.57861537952E-04)*3600 = 4.5573 PA: -2.55600299835E+01

%L1157-mm CARMA observations of the Full Stokes mode for E-array ($\sim$4.6$\arcsec$ resolution) and D-array ($\sim$2.0$\arcsec$) were conducted during July and September-October 2011 respectively. As part of the CARMA key polarimetric project called TADPOL (Telescope Array Doing POLarimetry, e.g., \citealt{hul13}), we also conducted C-array observations ($\sim$0.8$\arcsec$) in March-April 2012. An additional D-array track were received in June of 2012. Data reduction of all CARMA data used the MIRIAD software package (Sault et al. 1995).
%E, D, and C array observations were observed for X, Y, Z hours with a sensitivity of X, Y, Z Jy/bm respectively. %26 July, 29th september
CARMA observations of L1157 in full-Stokes mode at 1.3~mm were first made in the E-array ($\sim$4.6$\arcsec$ resolution) in July 2011 as a CARMA summer school project.  Additional observations in the D- ($\sim$2$\arcsec$) and C- ($\sim$0.8$\arcsec$) arrays were obtained over the next year as part of the TADPOL (Telescope Array Doing POLarization) key project \citep[e.g.,][]{hul13}. An additional D-array summer school observation was also added in June 2012. %Data reduction of all CARMA data used the MIRIAD software package (Sault et al. 1995).

The CARMA dual-polarization system is capable of measuring all 4 cross-correlations simultaneously through the circularly polarized feeds and full-Stokes correlator. The reduction process for the data in this paper is explained in \citet{hul13}. For L1157-mm, the flux calibrator was typically MWC349, but when this source was unavailable, we used Mars or Neptune. Flux calibration is accurate within $\sim$15\%; however, for the rest of the paper, only statistical uncertainties are discussed.  The phase calibrator used for all tracks was 1927+739.

\section{Results} \label{sec:results}
\begin{figure}[ht!]
\begin{center}
\includegraphics[scale=.7]{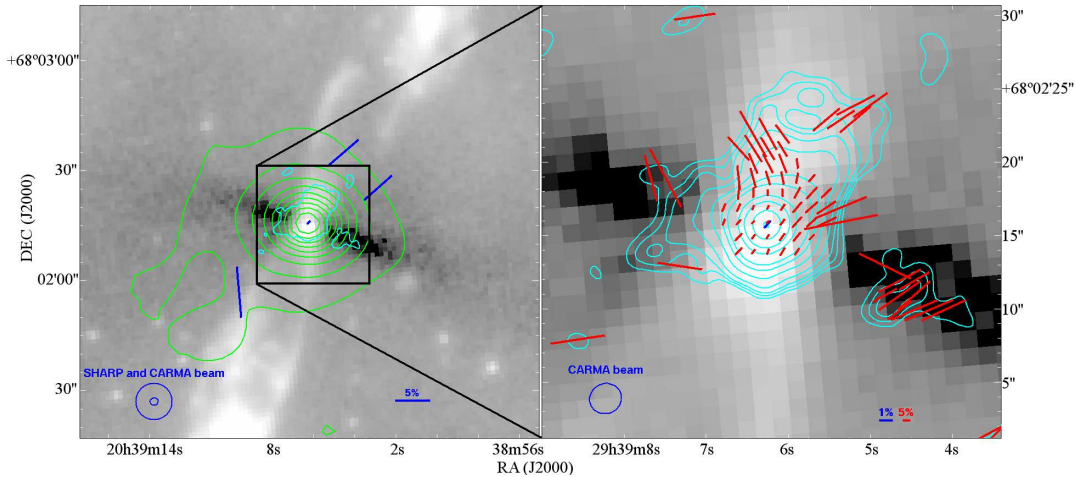}
\caption{Polarimetric maps (with polarization vectors rotated by 90$^\circ$ to show the inferred magnetic field orientation) of L1157-mm with the grayscale background showing a log-scale map of $Spitzer$ 8 $\mu$m emission \citep{loo07}. Magnetic field vectors are shown for $P_I>2\sigma_{P_I}$ and $I>2\sigma_I$. Left: SHARP continuum contours and vectors are shown in green and blue respectively. Cyan shows the 2$\sigma$ intensity detections from CARMA. The length of the vectors is proportional to P. SHARP contour levels range from 10\% to 90\% in 10\% increments of the peak flux. Right: Cyan (red) contours (vectors) are the combined CARMA data at 2.1$\arcsec$ resolution and the central blue vector is from SHARP. CARMA contour levels are [2,3,4,6,10,15,20,40,60,100,140]$\times \sigma$, $\sigma$=1.02 mJy/bm.  Negative contours are not shown. %in order to make the plot look more clean.
 \label{fig1.eps} 
}%The central polarization vector (where SHARP and CARMA overlap) is 0.7$\pm$0.2\% and 3.47$\pm$0.03\% for SHARP and CARMA, respectively.
\end{center}
\end{figure}

Figure \ref{fig1.eps} shows the results (dust continuum and the inferred magnetic field orientation) for SHARP and the combined CARMA observations. In the 350 $\mu$m SHARP map, an extension of the continuum is seen toward the southeast with a positional angle (PA, measured counterclockwise from north) of about 124$^\circ$, differing from the PA of the outflow of 161$^\circ$ \citep{bac01}; however, this extension roughly coincides with N$_2$H$^+$ extension seen in \citet{chi10}. In the combined CARMA data, the continuum is extended toward the northwest with PA~$\approx$~-20$^\circ$, which agrees well with $\sim$ -27$^\circ$ from $Spitzer$ 8$\mu$m and with -19$^\circ$ from \citet{bac01}. This extension may be due to photons preferentially escaping through the poles, thus heating the outflow cavity. The continuum also extends toward the east, showing the flattened envelope of L1157-mm; this extension matches very well with $Spitzer$ and N$_2$H$^+$ observations of the flattened structure \citep{chi10}. Additionally, the continuum observations show an indication of this flattened structure toward the west. The east and west continuum extensions are also seen at 1.3~mm with the SMA at $\sim$5.5$\arcsec$ resolution \citep{tob13}.

%SHARP (green contours and blue vectors) and the combined CARMA observations (cyan contours and red vectors). Magnetic field vectors are shown for $P_I>2\sigma_{P_I}$ and $I>2\sigma_I$.

% \begin{figure}[ht!]
% \begin{center}
% \includegraphics[scale=0.32]{all_and_zoom.eps}
% \caption{stuff
% \label{all_and_zoom.eps} 
% }
% \end{center}
% \end{figure}

%For SHARP, the central vectors and the two vectors in the northwest are somewhat aligned with the outflow; however, the southeast vector, though more aligned with the outflow ($<$45$^\circ$) than not, differs from the outflow. %It is possible that this vector is an anomaly (given $P$ = 2.1$\sigma_p$ and $\sigma_{\theta}$ = 10$^\circ$) and should be aligned with the outflow. However, this angle may signify that far from the source, the magnetic field about the outflow is misaligned.

%For SHARP, the central vector and the two vectors in the northwest are aligned within $\sim30^\circ$ with the outflow; however, the southeast vector is offset by XX degrees from the outflow (and XX degrees from the mean direction of the other three).

\begin{figure}[ht!]
\begin{center}
\includegraphics[scale=0.5]{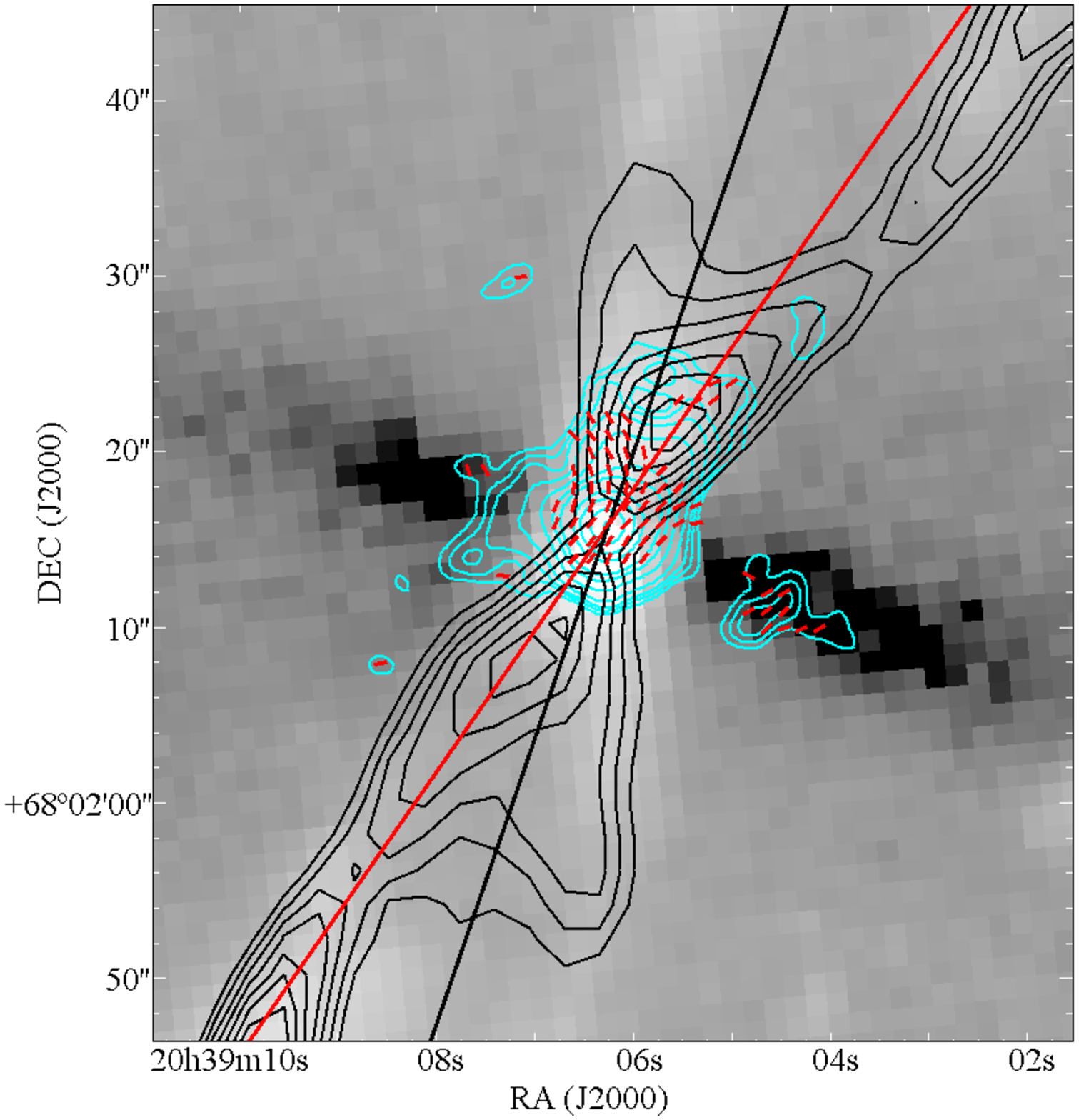}
\caption{Hourglass morphology of L1157 with the red line showing the axis of the hourglass and the black line showing the center of the outflow from \citet{bac01}. Cyan and red vectors are the same as Figure \ref{fig1.eps}. Black contours show the CO(2-1) moment-0 map (integrated intensity) of the outflow with contour levels of [2,3,4,5,6,7,8,9]$\times \sigma$, $\sigma$=8 Jy/beam km/s (Kwon et al. in preparation). Negative contours are not shown. % in order to make the plot look more clean.
\label{fig2.eps} 
}
\end{center}
\end{figure}

For SHARP, all vectors are aligned within $\sim$30$^\circ$ with the outflow angle measured by \cite{bac01};  however, the southeast vector is offset from the mean of the other three vectors by 50$^\circ$. CARMA vectors have a definitive hourglass shape, with the hourglass axis nearly coinciding with that of the outflow (Figure \ref{fig2.eps}).  We note that the difference between the hourglass axis and the outflow axis (a proxy for rotation axis) is about 15$^\circ$\footnote{\citet{hul13} derived a PA difference of 0$^\circ$ for L1157 because they defined the outflow direction as the line connecting the two brightest CO(2-1) peaks, whereas we follow \citet{bac01}, which uses the midline of the outflow cones.}. We do not discuss the reason for this discrepancy, though we note that Kwon et al. (in preparation) discusses the possibility of two jets ejected from L1157-mm, and our hourglass is tilted toward the more CO-bright (and probably more massive) jet. We note that the northwest SHARP vectors also coincides better with the CO-bright jet (within $\sim$15$^\circ$) than the entire outflow (within $\sim$30$^\circ$).

%Figure \ref{fig2.eps} shows the hourglass morphology with the red line showing the approximate axis of the hourglass and the black line showing the PA of the outflow from \citet{bac01}. Black contours show the integrated intensity CO(2-1) map from ancillary data (the entire outflow is mapped at 4$\arcsec$ resolution by Kwon et al. in preparation), and we note that the angle between the black and red lines is about 15$^\circ$\footnote{\citet{hul13} derived a PA difference of 0$^\circ$ for L1157 because they defined the outflow direction as the line connecting the two brightest CO(2-1) peaks, whereas we follow \citet{bac01}, which uses the midline of the outflow cones.}. We do not discuss the reason for this discrepancy, though we note that Kwon et al. (in preparation) discusses the possibility of two jets ejected from L1157-mm, and our hourglass is tilted toward the more CO-bright (and probably more massive) jet. We note that the northwest SHARP vectors also coincides better with the CO-bright jet (within $\sim$15$^\circ$) than the entire outflow (within $\sim$30$^\circ$).

%It is also a possibility that the northern polarization vectors are affected to some degree by the northern outflow lobe, since emission from this region is consistent with the emission of dust on the outflow cavity walls.  In such a region, the magnetic fields may be disturbed by the outflow; such distortions of other Class 0 sources have been suspected \citep{dav11}.

As seen in Figure \ref{fig1.eps}, we also find that for both CARMA and SHARP observations, $P$ is significantly less toward the center of the object, which is typical in polarimetric observations \citep[e.g.,][]{gir06}. Several factors may contribute to lower central fractional polarization, such as: (1) averaging along the line of sight is more likely to smear out the polarization through the thickest part of the source, (2) de-polarization at higher density, and (3) different grain populations at the center. 

\subsection{Comparison between different wavelengths}
%The central vectors have magnetic-field vectors of $P$=$0.7\pm0.2$\%, $\theta$=$-39.5\pm9.0^\circ$ for SHARP and P=$3.29\pm0.26$\%, $\theta$=$-35.1\pm2.3^\circ$ for the CARMA data. These vectors are our only point of comparison, and the angles are consistent with each other. This central angle is also consistent at all CARMA resolutions (see Figure \ref{fig3.eps}).
To compare SHARP and CARMA values, we smooth the CARMA data to a beam size of 10$\arcsec$  to provide accurate comparison between wavelengths. We find that the central vectors have  $P$=$0.7\pm0.2$\%, $\theta$=$-37.9\pm9.0^\circ$ for SHARP and P=$3.80\pm0.11$\%, $\theta$=$-32.2\pm0.8^\circ$ for the CARMA data. These vectors are our only point of comparison, and the angles are consistent with each other. This central angle is also consistent at all CARMA resolutions (see Figure \ref{fig3.eps}).

%9'': P: 3.944E-02 P_e=1.134E-03 PA: -31.88  (0.8 error)
%10'': P: 3.797E-02 P_e=1.092E-03 PA: -32.22 PA_e: 8.197E-01

%The fractional polarization of the SHARP and CARMA central vectors differs greatly, with $P$[1300]/$P$[350] $\approx$ 5.6. However, comparisons of single dish and interferometric fractional polarization should be made with caution. There are at least two competing effects: (1) higher resolution observations focus more on the core where $P$ typically is reduced; and (2) with varying polarization across the source, the interferometer tends to resolve out Stokes $I$ more than Stokes $Q$ or $U$, causing $P$ to be overestimated.

Studies have attempted to explain polarization spectra observed in various molecular clouds. \citet{vai12} present $P[\lambda]/P[350]$ values at wavelengths, $\lambda$, of 60, 100, 850, and 1300 $\mu$m. The median $P[\lambda]/P[350]$ value of a cloud can vary dramatically, but the typical value for these wavelengths was $\sim$2. Our value of $P$[1300]/$P$[350] $\approx$ 5.4 does not fit well within this polarization spectrum. However, comparisons of single dish and interferometric fractional polarization should be made with caution; e.g., with varying polarization across the source, the interferometer tends to resolve out Stokes $I$ more than Stokes $Q$ or $U$, causing $P$ to be overestimated. Additionally, we only have one point of comparison.

\subsection{Comparison between different size scales}
%The size-scale of where we see the hourglass shape is an important diagnostic to observationally access because low-mass star formation models \citep{fie93,gal93}\footnote{Or so says Girart06. I should carefully read these papers to confirm} predicts such a morphology at scales on order of a few 100 AUs. 
When a cloud with initially parallel magnetic field lines collapses and flux-freezing holds at least partially, the field becomes pinched into an hourglass morphology. According to ambipolar diffusion low-mass star formation models \citep[e.g.,][]{fie93,gal93,all03}, the hourglass size may range from the size of the protostar cloud core (a few hundred to a few thousand AU) to the size of the protostar infall envelope (up to about 10,000 AU). Our observations probe resolutions from $\sim$300-2250 AU and will provide a diagnostic of the size scale of the pinch.

\begin{figure}[ht!]
\begin{center}
\includegraphics[scale=0.65]{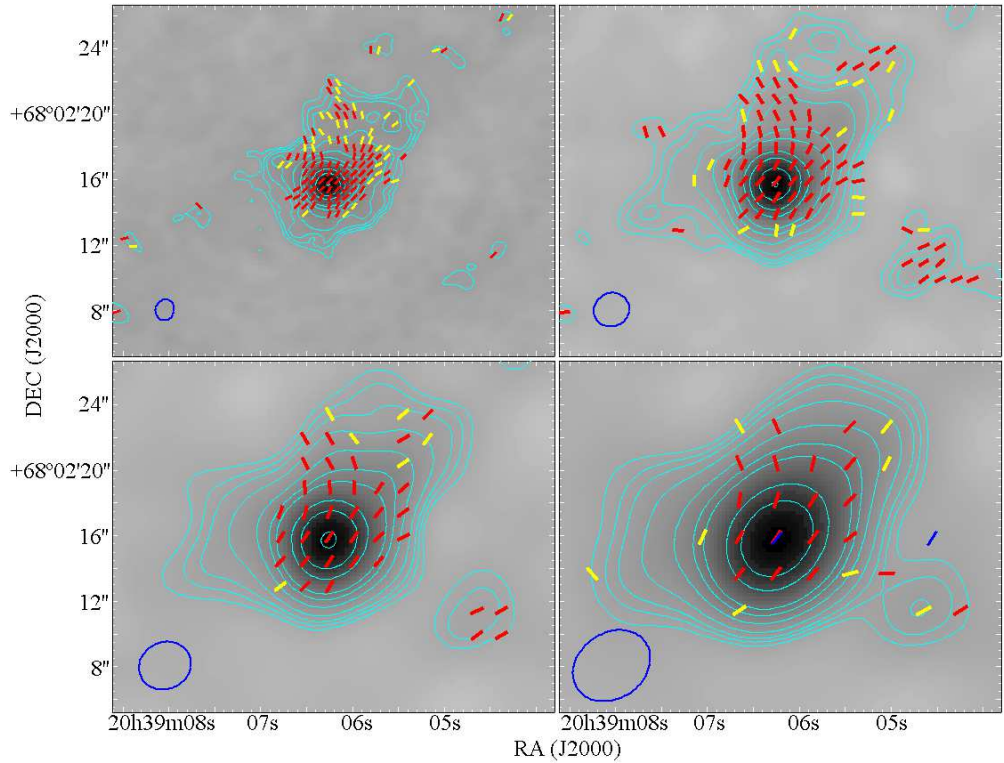}
\caption{The dust continuum emission and magnetic field morphology as measured by CARMA to determine the size-scale at which we see the hourglass. Grayscale is the Stokes I intensity on a square root scale and the blue ellipses show the beam size for each resolution. All vectors have $I>2\sigma_I$, the red vectors have $P_I>2\sigma_{P_I}$, and the yellow vectors have 1.5$\sigma_{P_I}<P_I<2\sigma_{P_I}$.  \emph{Top Left}: 1.2$\arcsec$ resolution, \emph{Top Right}: 2.1$\arcsec$ resolution, \emph{Bottom Left}: 3.0$\arcsec$ resolution, and \emph{Bottom Right}: 4.5$\arcsec$ resolution, with SHARP vectors for $P_I>1.5\sigma_{P_I}$ in blue (9$\arcsec$ resolution). The hourglass pinch becomes the most obvious at higher resolution. CARMA contour levels are the same as in Figure 1. The rms error, $\sigma$, for each of the four panels is 0.94, 1.02, 1.79, and 2.58 mJy/bm respectively. % in order to make the plot look more clean.
\label{fig3.eps} 
}
\end{center}
\end{figure}

From our combined CARMA observations, we apply 3 different Gaussian tapers to the visibility data which, along with the non-tapered data, probe 4 different resolutions (1.2$\arcsec$, 2.1$\arcsec$, 3.0$\arcsec$, and 4.5$\arcsec$) as seen in Figure \ref{fig3.eps}. We note that L1157-mm is only marginally resolved at 4.5$\arcsec$ resolution. Red vectors are shown when $P_I>2\sigma_{P_I}$ and are shown spatially at approximately Nyquist frequency ($\sim$2 vectors per beam). In attempt to see the most morphology possible given our sensitivity, we also show yellow vectors for 1.5$\sigma_{P_I}<P_I<2\sigma_{P_I}$.\footnote{When plotting vectors even down to $P_I$/$\sigma_{P_I}$=1, we notice that the scatter from the morphology is typically small ($\lesssim$10$^{\circ}$, indicating we have overestimated errors and/or vectors are not independent). Therefore, 1.5$\sigma_{P_I}$ vectors help in displaying the magnetic field morphology.}

%NO TRUE
%sig_theta = 1 radian/(1/2*S/N) ??
%Smaller $\sigma$ values on the de-biased $P_I$ are consistent with 0 (see discussion in \citealt{vai12}). \emph{Note: this is based off $P$, not $P_I$. Double-check the math eventually}

% P$>$1.28$\sigma_P$ (i.e., 90\% confidence that it is a significant detection; this limit should be change, see /citet{vai12})
%Observations in each array do not have the sensitivity to accurately show the polarization vectors throughout the entire region. Thus to predict this size-scale, we plot the polarization vectors of all 4 observations (SHARP and the 3 different CARMA array configurations) at lower significance. Figure \ref{fig3.eps} shows polarization vectors significant at P$_I > 1.44\sigma_{P_I}$ (92.5\% confidence P$_I > \sigma_{P_I}$) and $I > 2\sigma_I$ for SHARP and the three different array configurations used for observations with CARMA (C, D, and E in order of decreasing resolution) and the combined CARMA observations. Though most vectors have much higher significance than 1.44$\sigma_{P_I}$, these vectors imply a $\sigma_{\theta}$ = 20$^\circ$.
% Average $Q$ and $U$ noises for C-, D-, and E-array were 0.38, 0.53, 2.52 mJy/bm respectively. Even though the E-array has by far the worst sensitivity, we find that it still has the most detections, indicating that we resolve out emission in the C- and D-array configurations.

Parts of the hourglass morphology can be detected in the lower resolution maps, but the full hourglass becomes evident in the two higher resolution plots.  SHARP data, however, fail to see a definitive pinch even when considering lower $\sigma$ detections (the entire low-$\sigma$ SHARP polarization field is not shown in this letter).

The resolution at which the $full$ hourglass morphology becomes apparent at 1.3~mm is approximately 2.1$\arcsec$ or 550~AU. We note that this size-scale is highly uncertain (particularly due to distance and signal to noise), though it is roughly consistent with the 290~AU resolution probing the hourglass in \citet{gir06}, assuming a distance to NGC~1333 of 235~pc \citep{loi12}.

%This scale is consistent with the ambipolar diffusion low-mass star formation models \citep{fie93,gal93} and the size-scale probing the hourglass in \citet{gir06} of 360~AU. 

%The E-array certainly detects this pinch at the size-scale of $\sim$1150~AU, but it is the most prevalent at the D-array at 500~AU. It is not, however, detected at size scales of 2250~AU with SHARP. These scales are consistent with the ambipolar diffusion low-mass star formation models \citep{fie93,gal93} and the size-scale probing the hourglass in \citet{gir06} of 360~AU. 

\subsection{Magnetic Field Strength}
The strength of the magnetic field cannot be measured directly from dust polarization maps; however, different methods allow for estimations. The standard technique for calculating magnetic fields from polarimetric maps is the Chandraeskar-Fermi (CF) technique \citep{cha53} and modifications of the method. %\citep[e.g.,][]{ost01}. 

We use a modified CF technique \citep{ost01} in the large central region where polarimetric observations exist ($P_I>2\sigma_{P_I}$ and $I>2\sigma_I$ for the combined data at 2.1$\arcsec$ resolution). Our application of the technique follows the methodology outlined in \citet{gir06}, which fits parabolas to the vectors and uses the residuals from the fits to calculate the angle dispersion in NGC~1333 IRAS 4A. 

\begin{figure}[ht!]
\begin{center}
\includegraphics[scale=0.4]{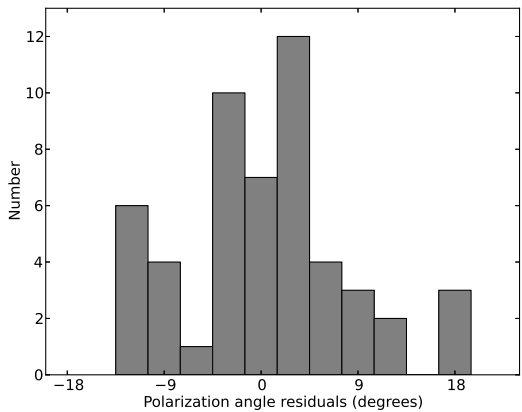}
\caption{Histogram of angle residuals (difference between measured angle and parabolic fit) at each vector location. One vector in the west of L1157-mm was excluded from the fitting since it caused a large reduced $\chi^2$ value.
%The outlier at $\sim$33$^\circ$ was not used in the Gaussian fit to calculate the dispersion since it significantly alters the residual mean from 0$^\circ$.
\label{fig4.eps} 
}
\end{center}
\end{figure}

The central region of L1157-mm has a total flux of 0.35 Jy in an area of 70 square arcsec. Using a dust temperature of 25~K (Chiang, private communication) and a dust opacity of  $\kappa_\nu$=0.9~cm$^2$/g \citep{oss94}, we find that the central region has a mean volume density of $n$(H$_2$)=7.4$\times 10^6$~cm$^{-3}$. From N$_2$H$^+$ observations, we use a non-thermal rms velocity dispersion about the line of sight of $\delta$v$_{los}$=0.18~km/s \citep{chi10}.  The dispersion from the residuals of parabola fits was 7.5$^{\circ}$ (see Figure \ref{fig4.eps}) and the uncertainty of the polarization angle is $\sim$4$^\circ$, resulting in an angular dispersion of $\delta \phi$=6$^{\circ}$. Given these parameters, we find a plane-of-sky magnetic field measurement of 1.4~mG.  Using these values, the mass-to-magnetic flux ratio \citep[using the method discussed in][]{cru99} is found to be 1.1 times the critical value for collapse. 

%. IRAS 4A and L1157-mm have similar velocity dispersions \citep{dif01,chi10}, so we adopt the same kinematically corrected velocity dispersion about the line of sight as \citet{gir06} of $\delta$v$_{los}$=0.2~km/s. 

Since the CF method only calculates the average magnetic field strength and has numerous uncertainties \citep[see discussion in][]{cru12}, we also use another method by \citet{koc12} that calculates the magnetic field strength at all locations where there are dust polarization and emission measurements. This technique assumes ideal magnetohydrodynamics (MHD) and that the dust emission gradient indicates the direction of the MHD force equation, leading to the equation for the magnetic field strength below:

\begin{equation}\label{eq:B_field}
B = \sqrt{\frac{\mbox{sin}~\psi}{\mbox{sin}~\alpha}(\nabla P + \rho \nabla \phi)~4 \pi R}
\end{equation}

In this equation, $\psi$ is the angle difference between gravity and the intensity gradient, $\alpha$ is the difference between polarization and the intensity gradient, P is the hydrostatic pressure, $\phi$ is the gravitational potential as a function of radius, and R is the local curvature radius of the polarization vectors. When applying this technique, we assume a density profile of the envelope to follow the fits of L1157-mm observations in \citet{chi12} (power-law of $\rho \propto r^{-p}$ with $p=2$). We also use a central core mass of M$_0$ = 0.19~M$_\sun$ \citep[assumed to be similar to L1527,][]{tob12}. Additionally, we assume the pressure gradient, $\nabla$P, to be negligible compared to gravity. The local curvature is calculated in the manner suggested in \citet{koc12}.

\begin{figure}[ht!]
\begin{center}
\includegraphics[scale=0.4]{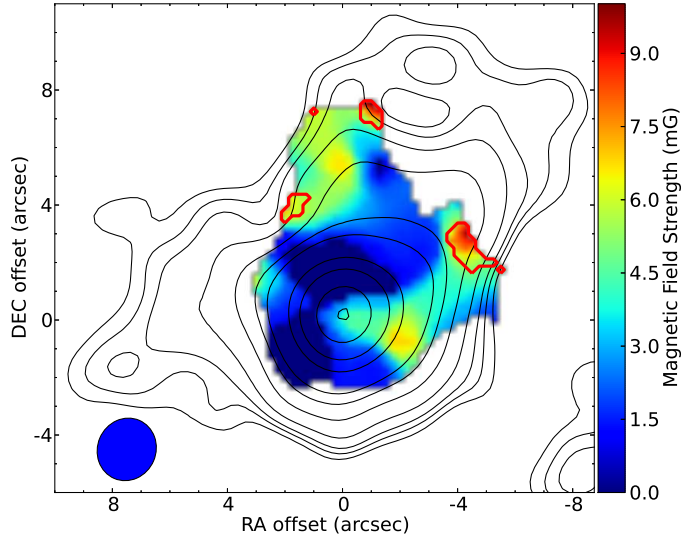}
\caption{Magnetic field throughout L1157-mm with the color scale in mG. Black contours are the same as Figure 1. Red contours indicate subcritical locations, while the rest of the colored area is supercritical. 
\label{fig5.eps} 
}
\end{center}
\end{figure}

The magnetic field throughout the large central region where polarimetric observations exist ($P_I>2\sigma_{P_I}$ and $I>2\sigma_I$ for the combined data at 2.1$\arcsec$ resolution) can be seen in Figure \ref{fig5.eps}; outside this area, magnetic field measurements are unreliable. This figure uses median-filtered smoothing at approximately Nyquist sampling (i.e., with pixel sizes of 0.25$\arcsec$, we replace each pixel value by the median of the surrounding 5$\times$5 box).  The colorbar indicates the field strength in mG, with an average and median magnetic field throughout the region of 3.6 and 3.4 mG respectively. The red contours show the locations where magnetic field tension dominates gravity (subcritical). Note that the central areas are supercritical (gravity dominates the magnetic field tension). The \citet{koc12} method determines the criticality based on $\mbox{sin}~\psi/\mbox{sin}~\alpha$, while the critical values for the CF method is an average about the entire region and depends on calculated values (e.g., magnetic field strength, mass).

%Moreover, it can also be seen that the magnetic field seems to decrease in strength from outside to inside, which is contrary to an ambipolar diffusion model. The magnetic field strengths in the northwest extension of the core influenced by the emission from the outflow cavity could be underestimated, since the fields in this region of the map are distorted by the outflow as well as gravity.

%The magnetic field is dependent on the uncertainty in the polarization angle, which depends on the accuracy in $P_I$. If we use only observations to $P_I>3\sigma_{P_I}$, this method calculates a mean and median of 3.0 and 2.2 mG respectively.
%2sig
%0.00425061223204
%0.0029991068345

%3sig
%0.00297769080094
%0.002213876755

When using the method from \citet{koc12} to estimate the magnetic field, we have 3 free parameters: $p$, $M_0$, and distance, $d$. In the parameter space of our uncertainty in variables, $1.5 < p < 2.5$, 0.1~M$_{\sun} < M_0 < 0.75$~M$_{\sun}$, and 200~pc $< d <$ 500~pc, the median magnetic field follows the approximate relationship:
\begin{equation}
B = 3.4~\mbox{mG} \left(\frac{p}{2}\right)^{1.9}\left(\frac{M_0}{0.19~\mbox{M$_{\sun}$}}\right)^{0.5}\left(\frac{d}{250~\mbox{pc}}\right)^{0.5}
\end{equation}

When altering these parameters, our median magnetic field value of 3.4~mG does not drastically change and will generally be within a factor of $\sim$2. Note that changing these parameters does not change the supercritical and subcritical locations in the cloud since these locations only depend on the value of $\mbox{sin}~\psi/\mbox{sin}~\alpha$.

% \begin{equation}
% \rho = \rho_0 (\frac{r}{r_0})^{-p}
% \end{equation}
% 

\section{Discussion}
In this letter we present the first detection of magnetic fields in L1157-mm at two different wavelengths. Our main results are:
\begin{itemize}
\item We find the second instance of a full hourglass morphology in the inferred magnetic field around a low-mass protostar. This is also only the second instance around a Class 0 source.  The axis of the hourglass is nearly aligned with the axis of the outflow, with the full detection of the hourglass occurring at a resolution of $\sim$550 AU.
\item The angle of the central magnetic field vector agrees at 350~$\mu$m and 1.3~mm and at all size-scales.
\item We used two methods to calculate the plane-of-sky magnetic field throughout the central region and find values of 1.4 and 3.4~mG.
%The magnitude of the plane-of-sky magnetic field of the central region its calculated by two methods are 1.4 and 3.4~mG.
%\item The L1157-mm map of the dust continuum traces both the outflow and the flattened envelope.
%\item The core of L1157 is about 5 times more polarized at 1 mm than 350 $\mu$m.
\end{itemize}

%Although it is precarious to compare $P$ with single dish and interferometric observations, we also find that the center of L1157-mm is about 5.6 times more polarized at 1.3~mm than at 350 $\mu$m.

%This hourglass morphology is analogous to the one seen in NGC 1333 IRAS 4A, a similar (though a definite binary) system as L1157-mm \citep{gir06,att09}. They also found an hourglass shape that was not completely symmetric, but their hourglass is not centered about the primary outflow.  This observation of L1157-mm is arguably the best example of an hourglass shape about a Class 0 source to date. Recent high resolution Expanded Very Large Array observations detect only a single continuum source at size-scales of $\sim$12~AU (Tobin, private communication), suggesting that L1157-mm is likely a single source. Perhaps a single system increases the likelihood of an hourglass detection and/or that the hourglass axis is centered about the outflow.

While this particular hourglass is nearly aligned with the outflow, the TADPOL survey results have shown that in general, magnetic field lines are consistent with being preferentially misaligned (perpendicular) or randomly aligned with respect to outflows \citep{hul13}.

The first well-defined hourglass morphology around a low-mass protostar was observed in 1997 in the Class 0 binary system NGC 1333 IRAS 4A \citep{gir99}; such a discovery has not been published since. Our observation of L1157-mm is arguably the best example of an hourglass shape in any star formation region to date. Characterizing such systems is an important step to better understand the role of magnetic fields in star formation.

IRAS 4A's hourglass axis of symmetry is misaligned with large scale outflow as measured by CO(3-2) \citep{bla95} by 16$^\circ$ and is over 40$^\circ$ misaligned with the small scale outflow as measured by SiO(1-0) \citep{cho05}. However, SHARP observations found that the large scale field in IRAS 4A coincides within 1$^\circ$ of the large scale outflow \citep{att09}. Our observations of L1157-mm find that the hourglass axis of symmetry and northwest SHARP vectors are similarly misaligned by $\sim$15-30$^\circ$, but the hourglass axis is more aligned with the outflow than these SHARP vectors. These results suggest the need to compare more polarimetric observations at different resolutions, which will be a future TADPOL project.

Recent high resolution Expanded Very Large Array observations detect only a single continuum source at size scales of $\sim$12 AU (Tobin, private communication), suggesting that L1157-mm is most likely a single source. Perhaps cores with a single system are more likely to yield an hourglass detection and/or an hourglass axis aligned with the outflow.

\acknowledgments
We thank Katherine Rosenfeld, Che-Yu Chen, and Aaron Juarez for their help in providing CARMA summer school tracks.

Support for CARMA construction was derived from the states of California, Illinois, and Maryland, the James S. McDonnell Foundation, the Gordon and Betty Moore Foundation, the Kenneth T. and Eileen L. Norris Foundation, the University of Chicago, the Associates of the California Institute of Technology, and the National Science Foundation. Ongoing CARMA development and operations are supported by the National Science Foundation under a cooperative agreement (NSF AST 08-38226) and by the CARMA partner universities. 

The Caltech Submillimeter Observatory is operated by the California Institute of Technology under cooperative agreement with the National Science Foundation (AST-0838261), and SHARP is supported by NSF grant AST-090930 to Northwestern University.

\end{document}